\documentclass[aps,prr,twocolumn,letterpaper]{revtex4-1}
\usepackage{hyperref}
\usepackage{amssymb,amsmath,multirow,color,natbib,bm,float}

\usepackage[T1]{fontenc}

\usepackage{mathptmx}
\usepackage{easyReview}
\newcommand{\pder}[2]{\frac{\partial{#1}}{\partial{#2}}}

\renewcommand{\vec}[1]{\mathbf{#1}}

\newcommand{\ez}{\vec{\hat e}_z}

\newcommand{\eref}[1]{(\ref{#1})}

\newcommand{\etal}{\emph{et.al.}}

\newcommand{\aave}[1]{\left\langle #1\right\rangle} 

\renewcommand{\vec}[1]{\mathbf{#1}}

\newcommand{\dk}{d\vec k}
\renewcommand{\d}[1]{d\vec{#1}}

\newcommand{\ee}[1]{{e}^{#1}}
\newcommand{\dkpq}[3]{\delta_{{#1},{#2}{#3}}}
\newcommand{\dkp}[2]{\delta(\vec{#1}+\vec{#2})}
\newcommand{\nlt}[3]{\int d{\vec #2}d{\vec #3}M_{{#1},{#2}{#3}}}
\newcommand{\ppsi}[2]{\psi^s_{#1}(#2)}
\newcommand{\mpsi}[2]{\psi^{-s}_{#1}(#2)}

\newcommand{\tgam}{\tilde\Gamma}
\newcommand{\mkpq}[3]{M_{{#1},{#2}{#3}}}
\newcommand{\va}{v_A}

\begin{document}

\author{Jean C. Perez}
\author{Augustus A. Azelis}
\author{Sofiane Bourouaine}
\affiliation{Florida Institute of Technology, 150 W University Blvd, Melbourne, Florida, 32901, USA}

\title{Two-time energy spectrum of weak magnetohydrodynamic turbulence}

\begin{abstract}
  In this work a weak-turbulence closure is used to determine the structure of the two-time power spectrum of weak magnetohydrodynamic (MHD) turbulence from the nonlinear equations describing the dynamics. The two-time energy spectrum is a fundamental quantity in turbulence theory from which most statistical properties of a homogeneous turbulent system can be derived. A closely related quantity, obtained via a spatial Fourier transform, is the two-point two-time correlation function describing the space-time correlations arising from the underlying dynamics of the turbulent fluctuations. Both quantities are central in fundamental turbulence theories as well as in the analysis of turbulence experiments and simulations. However, a first-principles derivation of these quantities has remained elusive due to the statistical closure problem, in which dynamical equations for correlations at order $n$ depend on correlations of order $n+1$.  The recent launch of the \emph{Parker Solar Probe} (\emph{PSP}), which will explore the near-Sun region where the solar wind is born, has renewed the interest in the scientific community to understand the structure, and possible universal properties of space-time correlations. The weak MHD turbulence regime that we consider in this work allows for a natural asymptotic closure of the two-time spectrum, which may be applicable to other weak turbulence regimes found in fluids and plasmas. An integro-differential equation for the scale-dependent temporal correlation function is derived for incompressible Alfv\'enic fluctuations whose nonlinear dynamics is described by the reduced MHD equations.
\end{abstract}

\maketitle

\section{Introduction} Incompressible magnetohydrodynamics (MHD) not only provides the simplest framework to describe plas\-ma turbulence in magnetized plasmas, but often provides an accurate description of the low frequency nonlinear dynamics at spatial scales much larger than any plasma microscale, such as particle gyroradii~\cite{biskamp03}. MHD equations describe the dynamics of fluctuations of plasma velocity $\vec v(\vec x,t)$ and magnetic field $\vec B(\vec x,t)$ propagating along a background magnetic field $\vec B_0$. However, MHD turbulence is best studied in terms of the so-called Elsasser variables $\vec z^\pm=\vec v\pm\vec b$
\begin{equation} 
\left(\frac{\partial}{\partial t}-s\vec v_A\cdot\nabla\right)\vec z^s+\left(\vec z^{-s}\cdot\nabla\right)\vec z^s = -\nabla P,\label{mhd_elsasser}
\end{equation}
where $s=\pm$, ${\vec v}_A={\vec B}_0/\sqrt{4\pi \rho_0}$ is the Alfv\'en velocity associated with the constant background magnetic field ${\vec B}_0$, $\vec b=\vec B/\sqrt{4 \pi \rho_0}$ is the fluctuating Alfv\'en velocity, $P=(p/\rho_0+b^2/2)$, $p$ is the plasma pressure and $\rho_0$ is the background plasma density. Incompressibility imposes the additional condition $\nabla\cdot\vec z^s=0$. The advantage of the Elsasser description stems from the fact that nonlinear interactions responsible for the turbulence take place only when fields $\vec z^+$ and $\vec z^-$ spatially overlap, otherwise, $\vec z^s$ behave as noninteracting, nondispersive counterpropagating Alfv\'en waves.

Most of the recent theoretical and numerical progress in MHD turbulence so far has been concerned with understanding the spectral distribution of energy among spatial scales at a fixed time\citep{biskamp03,galtier00,lithwick03,lithwick07,boldyrev05,boldyrev09,mason06,perez08,perez09,perez12,chandran08,chandran15,mallet17a,boldyrev17}. Moreover, solar wind turbulent signals are used to investigate the spatial properties of the turbulent solar wind by relying on Taylor's hypothesis (TH)~\cite{taylor38},  which posits that the temporal variation of spacecraft signals is solely due to the spatial variation of frozen structures passing by the observation point. In contrast, the actual space-time structure of the turbulence has been investigated to a much lesser extent, perhaps in part because most testable theoretical predictions and models are concerned with its spatial structure, and because comparison with observations strongly relies on the TH~\cite{chen16}.  However, it has been recently shown that TH might not be valid in the near-sun region where the \emph{Parker Solar Probe} (\emph{PSP}) mission~\cite{fox16} will explore, and that the accurate interpretation of \emph{PSP} time signals will require an  understanding of how turbulent structures decorrelate in time~\cite{bourouaine18,chhiber19}.  In other words, modeling the two-point two-time  correlation function of MHD turbulence becomes a necessity for the interpretation of single-probe turbulent signals in the \emph{PSP} era, motivating our interest in this work on the physics of temporal decorrelation of turbulent fluctuations. 

Although dynamical equations governing the evolution of these space-time correlations can be formally obtained from the dynamical equations of the turbulent fluctuations, they inevitably result in an under-determined system of equations, i.e., one in which there are always more unknowns than equations. Because this so-called closure problem requires additional \emph{ad hoc} assumptions about the undetermined quantities, a first-principles derivation of the space-time correlation has remained elusive even in the simpler case of incompressible Hydrodynamic turbulence. As a result, investigations of space-time correlations in MHD turbulence have been limited to numerical simulations and phenomenological modeling~\cite{servidio11,lugones16,bourouaine19,bourouaine20,narita17a}. For instance,~\citet{servidio11}~and~\citet{lugones16} investigated the space-time correlation of magnetic fluctuations in simulations of MHD turbulence, with and without a guide field, and suggested that the temporal decorrelation of fluctuations may be associated with hydrodynamic and magnetic sweeping and/or Alfv\'enic propagation depending on the degree of magnetization. Recently, a number of phenomenological models have been proposed to explain results from numerical simulations and to develop new methodologies to interpret solar wind observations beyond TH~\cite{bourouaine19,bourouaine20,narita17a}, by extending Kraichnan's sweeping hypothesis of hydrodynamics~\cite{kraichnan64}, which states that the temporal decorrelation of small-scale turbulent fluctuations is due to the random advection by large scale velocity fluctuations. 

For spatial correlations (two-point one-time), a successful asymptotic closure has been accomplished in the weak turbulence (WT) regime to leading order in the turbulent amplitudes and without any \emph{ad-hoc} assumptions~\cite{galtier00}. The main objective of this work is to obtain a similar WT closure for the space-time correlation function. Although there is no evidence yet for the existence of weak MHD turbulence in the solar wind, the results of this work can potentially be used to establish the existence of such a regime from observations of current multi-spacecraft missions, such as \emph{Cluster}~\cite{balogh01} and \emph{Magnetospheric Multiscale} (\emph{MMS})~\cite{burch16}, as well as future missions currently being proposed to investigate solar wind turbulence by multipoint measurements from an array of spacecraft~\cite{klein19,tenbarge19}.

Aside from its potential relevance for current and future solar wind missions, the WT regime allows for a systematic asymptotic closure that can shed new physical insights and/or provide new theoretical constraints on the temporal decorrelation mechanisms of MHD turbulence. Furthermore, our results may also be applicable to other regimes of fluid and plasma turbulence, such as surface gravity waves~\cite{dyachenko04}, compressible MHD waves~\cite{chandran05}, gravitational waves in the early universe~\cite{galtier17}, waves in rotating planetary flows~\cite{galtier14}, and non-Newtonian electrohydrodynamic flows~\cite{carbone11}. 

\section{Background}
\subsection{Governing equations} For a strong background magnetic field $\vec B_0=| \vec B_0| \ez$, with $|\vec B_0| \gg|\vec b|$, the universal properties of MHD turbulence can be accurately described by neglecting the field-parallel component, $\vec z^s_\|$, of the fluctuating fields (the pseudo-Alfv\'en fluctuations) that play a sub-dominant role in the turbulence dynamics (see~\cite{perez12} and references therein), i.e. we can set $\vec z^s_\|=0$ in Eq.~\eref{mhd_elsasser} to obtain the following governing equation for the Fourier amplitudes
\begin{equation}
  \left(\frac{\partial}{\partial t}-sik_\|v_A\right)\vec{z}^s(\vec k,t)=-\epsilon\left[\left(\vec{z}^{-s}\cdot\nabla\right)\vec{z}^s\right]_{\vec k}^{\rm nc}, \label{rmhd-vector}
\end{equation}
where $\vec{z}^s(\vec k,t) = (z_x^s(\vec k,t),z^s_y(\vec k,t),0)$, $k_\|$ is the component of the wavevector $\vec k$ along the background magnetic field and $\epsilon$ is a dimensionless ordering parameter that quantifies the strength of the fluctuation amplitudes~\footnote{The ordering parameter $\epsilon$ can set to one after perturbative expansions are obtained to the desired order.}. The right-hand side of Eq.~\eqref{rmhd-vector} represents the Fourier-transformed nonlinear terms, which have the form of a convolution of the Fourier-amplitudes with the pressure properly chosen to ensure the fluctuations remain noncompressive (nc).  It can be shown that this model is equivalent to the two-field reduced MHD (RMHD) model~\cite{kadomtsev74,strauss76}.

By virtue of the incompressibility condition $\vec k_\perp\cdot\vec{z}^s=0$, we can introduce two Elsasser potentials $\psi^s(\vec k,t)$ 
\begin{equation}
  \vec{z}^s(\vec k,t) = i\psi^s(\vec k,t)\ee{isk_\|\va t}\vec{\hat e_{k}}\label{fields-potential},
\end{equation}
where $\vec{\hat e_{\vec k}}\equiv\vec k_\perp\times\ez/k_\perp$. The complex exponential in this definition is introduced as an integrating factor to account for the linear dynamics, in which case the fields $\psi^s(\vec k,t)$ describe the true variation of the wave amplitudes due to the nonlinear dynamics. From Eq.~\eqref{rmhd-vector}, the governing equations for the shear-Alfv\'en Fourier amplitudes $\psi^s(\vec k,t)$ become
\begin{equation}
   \pder{}{t}\ppsi kt = \epsilon\nlt kpq\ppsi pt\mpsi qt \ee{-2isq_\|v_At}\dkpq kpq,\label{eq:rmhd2}
\end{equation}
where
\begin{eqnarray}
  &&M_{k,pq}\equiv \frac{(\vec k_\perp\cdot\vec p_\perp)(\vec k_\perp\times\vec q_\perp)_{\|}}{k_\perp p_\perp q_\perp},\\
  &&\dkpq kpq \equiv\delta(\vec k-\vec p-\vec q),\quad\hbox{and}\quad  \ppsi kt\equiv \psi^s(\vec k,t).
\end{eqnarray}

\subsection{The two-point two-time correlation} The space-time structure of MHD turbulence is investigated using the two-point two-time correlation function
\begin{equation}
  C^s(\vec x,\vec x';t,t')=\aave{\vec z^s(\vec x,t)\cdot\vec z^s({\vec x}',t')},\label{eq:Rs} 
\end{equation}
which measures the covariance between the values of each Elsasser variable $\vec z^s$ at two different locations separated by a distance $\vec r=\vec x'-\vec x$ and with a time delay (or lag) $\tau=t'-t$. In Eq.~\eqref{eq:Rs} $\aave{\cdots}$ is used to denote ensemble average over turbulence realizations. For homogeneous turbulence this correlation depends only on the relative vector $\vec r$ and can be written as
\begin{equation}
  C^s(\vec r;t,t')=\int\dk h^s(\vec k,t,t')e^{-i\vec k\cdot\vec r}
\end{equation}
where
\begin{equation}
  h^s(\vec k,t,t')\equiv\left\langle\vec z^{s}(-\vec k,t)\cdot\vec z^{s}(\vec k,t')\right\rangle,\label{eq:laggedE}
\end{equation}
is the two-time power spectrum and $\vec z^s(\vec k,t)$ are the Fourier amplitudes of the Elsasser fields. The two-time spectra defined in Eq.~\eqref{eq:laggedE} are related to the two-time spectra for Elsasser potentials through the simple transformation
\begin{equation}
  h^s(\vec k,t,t')=\tilde h^s(\vec k,t,t')\ee{isk_\|\va(t'-t)},\label{eq:hkttp}
\end{equation}
where
\begin{equation}
\tilde h^s(\vec k,t,t')\equiv\aave{\ppsi{-k}t\ppsi k{t'}}.\label{eq:hkttp2}
\end{equation}

The space-time correlation function defined in Eq.~\eqref{eq:Rs} is not only one of the most fundamental quantities in turbulence theory but is also important in the analysis and interpretation of experimental and simulation data, such \emph{PSP} measurements in the spacecraft frame in terms of space-time properties in the plasma frame when the TH is no longer applicable~\cite{narita17a,bourouaine19,bourouaine20}. A better understanding of the space-time dynamics of the turbulence will allow observers to maximize the information and interpretation of \emph{PSP} measurements.

\subsection{The two-time vs one-time energy spectrum}
The one-time (spatial) three dimensional power spectra follows from the two-time spectra when $t'=t$,
\begin{equation}
  e^s(\vec k,t)\equiv\aave{\ppsi{-k}t\ppsi kt}=h^s(\vec k,t,t),\label{eq:eskt}
\end{equation}
which allows one to define the scale-dependent time correlation function as
\begin{equation}
  h^s(\vec k,t,t')=e^s(\vec k,t)\Gamma^s(\vec k,t,t'),\label{eq:two_vs_one0}
\end{equation}
where by definition $\Gamma^s_k(t,t)=1$. 

It is important to point out that two-time correlations and two-time power spectra capture additional information not present in their one-time counterparts. The former provide information about the time memory of the turbulence by relating turbulence properties at two separate times, while the latter contain only spatial information at any given time during the turbulence evolution.

A better interpretation of two-time correlations is normally provided through the simple change of variables $t,t'\rightarrow t,\tau=t'-t$. The role of each of these time variables in the two-time correlations is very different. On one hand, the time $t$ tracks the temporal evolution of the turbulence from some initial state at $t=t_0$, after which the turbulence can either decay in an undriven system or transition to a steady state in the driven case. On the other hand, the time variable $\tau$ simply represents a time delay introduced to investigate the time memory of the system at any given time $t$ through temporal correlations. Because fluctuations decorrelate in a finite time, it follows that $\Gamma^s(\vec k,\tau)\rightarrow 0$ when $\tau\rightarrow\infty$. Eq.~\eqref{eq:two_vs_one0} thus becomes
\begin{equation}
  h^s(\vec k,t,\tau)=e^s(\vec k,t)\Gamma^s(\vec k,t,\tau).\label{eq:two_vs_one}
\end{equation}
Weak turbulent closures for one-time quantities, in both fluids and plasmas~\cite{nazarenko11}, provide a closed wave-kinetic equation describing the true time evolution of the energy spectrum $e^s(\vec k,t)$ in the variable $t$. The wave-kinetic equation is thus an initial value problem from which decaying or steady state solutions can be obtained.

In the steady state, the one-time spectrum $e^s(\vec k)$ is independent of $t$ and two-time quantities are a function of $\tau$ only; therefore~\eqref{eq:two_vs_one} becomes
\begin{equation}
  h^s(\vec k,\tau)=e^s(\vec k)\Gamma^s(\vec k,\tau).\label{eq:hgamma}
\end{equation}

The two-time power spectrum can also be Fourier transformed with respect to the variable $\tau$ to define the wavevector-frequency power spectrum
\begin{equation}
  h^s(\vec k,\omega) = \frac 1{2\pi}\int h^s(\vec k,\tau)\ee{i\omega\tau}d\tau,\label{eq:hkomega}
\end{equation}
which measures the spectral distribution of energy by wavevector and frequency. 
Because the Fourier transform in this equation involves only $\tau$, it follows from Eq.~\eqref{eq:hgamma} that
\begin{equation}
  h^s(\vec k,\omega)=e^s(\vec k)\Gamma^s(\vec k,\omega),\label{eq:hgamma2}
\end{equation}
where $\Gamma^s(\vec k,\omega)$ is the Fourier transform of $\Gamma^s(\vec k,\tau)$ with respect to the variable $\tau$. The advantage of Eqs.~\eqref{eq:hgamma}~and~\eqref{eq:hgamma2} is that they allow for the separation of the purely spatial part of the correlation function, typically investigated in theory and simulations, from the scale-dependent temporal part, which is the subject of this work. The two quantities $h^s(\vec k,\tau)$ and $h^s(\vec k,\omega)$ are simply different representations of the correlation function $C^s(\vec r,\tau)$, obtained via Fourier transforms and thus contain the same information as the two-point two-time correlation function.

A large body of works have been devoted to understanding the structure of the two-time power spectrum, or equivalently the scale-dependent time correlations $\Gamma^s(\vec k,\tau)$ and corresponding Fourier transform $\Gamma^s(\vec k,\omega)$, from analytical closures, numerical simulations, and phenomenological modeling, in both hydrodynamics and MHD turbulence~\cite{servidio11,lugones16,bourouaine19,bourouaine20,narita17a,zhou04,zhou10}. For instance, based on purely heuristic arguments~\citet{zhou04} proposed the following general form for the scale-dependent time correlation of MHD turbulence
\begin{equation}
    \Gamma^s(\vec k,\tau)=\ee{isk_\|v_A\tau}\ee{-\gamma^s_{\rm nl}(k)\tau}\ee{-[\gamma^s_{\rm sw}(k)\tau]^2}.\label{eq:zhou}
\end{equation}
More recently phenomenological models have been proposed in the context of MHD turbulence, such as~\citet{narita17a} who extended Kraichnan's sweeping hypothesis to MHD turbulence to obtain a model of $\Gamma^s(\vec k,\omega)$ that is consistent with Eq.~\eqref{eq:zhou} for $\gamma_{\rm nl}=0$ and where the sweeping decorrelation  is $\gamma^s_{\rm sw}\propto z^{-s}_{\rm rms} k_\perp$, where $z^s_{\rm rms}$ represents the root mean square (rms) value of the $\vec z^s$ field at the outer scale. However,~\citet{bourouaine18} determined from simulations of imbalanced, reflection-driven MHD turbulence that the two Elsasser fields decorrelate at a common speed consistent with the rms of velocity at the outer scale, and later developed a new model for the correlation in which the sweeping is solely attributed to hydrodynamic sweeping, and $\gamma_{\rm sw}^s(k)\propto k_\perp u_0$~\cite{bourouaine19,bourouaine20}.

The main goal of this work is to obtain a closed equation that describes the $\tau$ dependence of the temporal correlations $h^s(\vec k,\tau)$, when the system is in steady state with respect to $t$, for the WT regime. The advantage of the WT turbulence regime is that a first-principles, albeit asymptotic, closure for the correlation can be obtained and compared with other models for the scale-dependent time correlations. In the next section we will first present a brief derivation of Galtier~\etal~\cite{galtier00} WT closure for $e^s(\vec k,t)$, which will later be extended to the two-time power $h^s(\vec k, \tau)$.

\section{WT closure for the one-time spectrum} 

In this section we present a concise derivation of~\citet{galtier00} WT closure for MHD turbulence with respect to the time $t$. For this purpose, Eqs.~\eqref{eq:rmhd2} and~\eqref{eq:eskt} are used to obtain the governing equation for the one-time spectrum $e^s_k(t)$ as follows
\begin{widetext}
\begin{eqnarray}
  \pder{}te^s_k(t)&=&\pder{}t\aave{\ppsi{-k}t\ppsi{k}{t}}=\aave{\pder{}t\ppsi
                                         {-k}t\ppsi kt}+\aave{\ppsi
                                         {-k}t\pder{}t\ppsi{k}t}\nonumber\\
                                     &=& \epsilon\int\d p\d q\left[M_{-k,pq}\aave{\ppsi
                                         {k}t\ppsi
                                         pt\mpsi qt}\dkpq{-k}pq+M_{k,pq}\aave{\ppsi
                                         {-k}t\ppsi
                                         pt\mpsi qt}\dkpq{k}pq\right]\ee{-2isq_\|v_At}.\label{eq:phi2}
\end{eqnarray}
\end{widetext}
The two terms in the last integral can be regrouped by changing $\vec p\rightarrow -\vec p$ and $\vec q\rightarrow\-\vec q$ in the first integral to get
\begin{eqnarray}
  \pder{}te^s_k(t)
  &=& 2\epsilon\nlt kpq{\rm Re}\left[Q^{s}_{-kp}(t)\ee{-2isq_\|v_At}\right]\dkpq{k}pq,\nonumber\\\label{eq:Second}
\end{eqnarray}
with the third-order correlation defined as
\begin{equation}
  \aave{\ppsi kt\ppsi p{t}\mpsi q{t}}\equiv Q_{kp}^{s}(t)\delta(\vec k+\vec p+\vec q)\label{eq:q3ot}.
\end{equation}
Similarly, the evolution of the third-order correlation
\begin{multline}
  \pder{}{t}\aave{\ppsi kt\ppsi p{t}\mpsi q{t}} = \\
  =\epsilon\nlt kln\aave{\ppsi lt\mpsi n{t}\ppsi p{t}\mpsi q{t}}\\
  \times\ee{-2isn_\|v_At}\dkpq kln+(sk\leftrightarrow sp\leftrightarrow -sq)\label{eq:Third}
\end{multline}
depends on fourth-order correlation functions. Here, $(sk\leftrightarrow sp\leftrightarrow -sq)$ indicates two more similar terms obtained from all unique permutations of  $sk,sp,-sq$. Fourth-order correlations can be rewritten in terms of the fourth-order cumulant, defined as
\begin{multline}
  \left\{\ppsi pt\ppsi l{t}\mpsi q{t}\mpsi n{t}\right\}=\aave{\ppsi pt\ppsi l{t}\mpsi q{t}\mpsi n{t}}\\-\aave{\ppsi pt\ppsi l{t}}\aave{\mpsi q{t}\mpsi n{t}}\\
  -\aave{\ppsi pt\mpsi q{t}}\aave{\ppsi l{t}\mpsi n{t}}\\
  -\aave{\ppsi pt\mpsi n{t}}\aave{\ppsi l{t}\mpsi q{t}}.
\end{multline}
Assuming negligible correlations between $\psi^s$ and $\psi^{-s}$ wave amplitudes it follows
\begin{multline}
  \aave{\ppsi pt\ppsi l{t}\mpsi q{t}\mpsi n{t}} = e^s_p(t)e^{-s}_q(t)\dkp pl\dkp qn\\
  +\left\{\ppsi pt\ppsi l{t}\mpsi q{t}\mpsi n{t}\right\}.\label{eq:Cumulant}
\end{multline}

The WT closure becomes possible, as can be rigorously shown~\cite{benney66,galtier00}, because for small $\epsilon$ the fourth-order cumulant evolves over two disparate timescales: a fast linear wave timescale at zero order $\epsilon^0t$ and a much slower nonlinear timescale at order $\epsilon^2t$ associated with the energy transfer between counterpropagating waves. Hence, to leading order in $\epsilon$, the contribution from the fourth-order cumulant to the evolution of $Q^{s}_{-kp}$ arises from the zero-order ($\epsilon^0$) wave amplitudes. However, at zero order the linear dynamics irreversibly drives the wave amplitudes towards a state of joint Gaussianity, in which case the fourth-order cumulant vanishes to zero-order in $\epsilon$. As rigorously shown by~\citet{benney66} for general systems of weakly interacting waves, the zero-order wave amplitudes can then be assumed to be Gaussian random fields in Eq.~\eqref{eq:Third}, thereby closing the moment hierarchy. Although nonlinear wave couplings regenerate fourth order cumulants over the longer $\epsilon^2t$ timescale, they do not contribute to leading order in Eq.~\eqref{eq:Third}.   Combining~\eqref{eq:q3ot}~\eqref{eq:Third} and~\eqref{eq:Cumulant} and discarding the fourth order cumulant we obtain a set of closed equations for the third-order moment
\begin{eqnarray}
  \pder{}t Q_{-kp}^{s}&=& \epsilon\mkpq kp{(k-p)}\left(e^s_p-e^s_k\right)e^{-s}_{k-p}\ee{2is(k_\|-p_\|)v_At}.\label{eq:Third-closed}
\end{eqnarray}

Long time solutions of Eq.~\eqref{eq:Third-closed} can be used in Eq.~\eqref{eq:Second} to obtain the wave-kinetic equation
\begin{equation}
  \pder{}te^s_k=\frac{\epsilon^2\pi}{v_A}\int d\vec pd\vec q\mkpq
  kpq^2\left(e^s_p-e^s_k\right)e^{-s}_q\delta(q_\|)\dkpq{k}pq.\label{eq:WKeqn}
\end{equation}

One important consequence of this closure is the fact that turbulent energy is strictly transferred to small field-perpendicular scales, given that three-wave interactions always involve modes with $q_\|=0$ and the resonance condition $\vec k=\vec p+\vec q$ implies $k_\|=p_\|$. As a consequence, $k_\|$--planes evolve independently and the parallel spatial structure of the turbulence is not altered. The power spectrum can then be factored out as $e^s(\mathbf k)=\mathcal E^s(k_\perp)g^s(k_\|)$, where $\mathcal E^s(k_\perp)$ is the field-perpendicular power spectrum for $\vec z^s$ waves defined so that
\begin{equation}
    E^s=\int dk_\perp\mathcal E^s(k_\perp)
\end{equation}
is the total energy of $\vec z^s$ waves. 
$g^s(k_\|)$ is a nonuniversal function that determines how the energy of the waves is distributed among the field-parallel scales~\footnote{It is worth stressing that this structure remains unchanged during the nonlinear cascade, so long as the turbulence remains weak.}.

\section{WT closure for the two-time spectrum} 
A similar procedure is followed to obtain a closure for the two-time correlation function, defined in Eq.~\eqref{eq:hkttp2}, to leading order in $\epsilon$. 
From the governing equations of $\ppsi kt$ it follows
\begin{multline}
  \pder{}{t'}\aave{\ppsi{-k}t\ppsi k{t'}}=\aave{\ppsi{-k}t\pder{}{t'}\ppsi{k}{t'}}\\
  =\epsilon\nlt{k}pqQ^{s}_{-kp}(t,t')
  \ee{-2isq_\|v_At'}\dkpq{k}pq\label{eq:second},
\end{multline}
where we again run into a similar problem in which the equations for second-order correlations depend on third-order ones, in this case at two different times,
\begin{equation}
  \aave{\ppsi {-k}t\ppsi p{t'}\mpsi q{t'}}\equiv Q_{-kp}^{s}(t,t')\delta(\vec k-\vec p-\vec q).
\end{equation}
In turn, the evolution of the third-order correlation is
\begin{multline}
  \pder{}{t'}\aave{\psi_{-k}^s{\psi'}_p^s{\psi'}_q^{-s}} \\
  = \epsilon\nlt pln\aave{\psi_{-k}^s{\psi'}_l^s{\psi'}_n^{-s}{\psi'}_q^{-s}}\ee{-2isn_\|v_At'}\dkpq pln\\
  +\epsilon\nlt qln\aave{\psi_{-k}^s{\psi'}_p^s{\psi'}_l^{-s}{\psi'}_n^s}\ee{2isn_\|v_At'}\dkpq qln\label{eq:third}
\end{multline}
where $\psi'$ is a shorthand for amplitudes evaluated at $t'$. As expected, the third-order correlation depends also on two-time fourth order correlations. Expressing the fourth order correlations in terms of second-order ones and fourth order cumulants~\footnote{The last term in Eq.~\eqref{eq:third} vanishes under the assumption of negligible correlations between counterpropagating fluctuations, as it was done for the WT closure of the one-time spectra.}
\begin{multline}
    \aave{\psi_{-k}^s{\psi'}_l^s{\psi'}_n^{-s}{\psi'}_q^{-s}}=\tilde h^s_{k}(t,t')\tilde h^{-s}_q(t',t')\delta(\vec k-\vec l)\dkp qn\\
    +\left\{\psi_{-k}^s{\psi'}_l^s{\psi'}_n^{-s}{\psi'}_q^{-s}\right\}.
\end{multline}
The fourth order cumulants in the last equation can also be discarded from Eq.~\eqref{eq:third} for long times (of order $\epsilon^2t$) for the same reason they are discarded in the one-time case, namely, that the leading contribution to the right-hand side of Eq.~\eqref{eq:third} arises from zero-order wave amplitudes, which reach a state of joint Gaussianity over the linear timescale of order $\epsilon^0t$. Therefore it follows from Eq.~\eqref{eq:third} that for long $t$
\begin{equation}
   \pder{}{t'}Q_{-kp}^{s}(t,t') =
   -\epsilon\mkpq{k}{p}{q}\tilde h^s_{k}(t,t')e^{-s}_q(t')\ee{2isq_\|v_At'}.\label{eq:q3}
\end{equation}
Eq.~\eqref{eq:second} and \eqref{eq:q3} now become a closed set of equations for the second- and third-order correlations. Eq.~\eqref{eq:q3} can be integrated between $t$ and $t'$ to obtain
\begin{multline}
  Q_{-kp}^{s}(t,t') = -\epsilon\mkpq{k}{p}{q}\int_t^{t'}dt''\tilde h^s_{-k}(t,t'')e^{-s}_q(t'')\ee{2isq_\|v_At''},
\end{multline}
which upon substitution in Eq.~\eqref{eq:second} leads to
\begin{multline}
  \pder{}{t'}\tilde h^s_k(t,t')=-\epsilon^2\int d\vec pd\vec q\mkpq kpq^2\dkpq kpq\\
  \times\int_t^{t'}dt''\tilde h^s_k(t,t'')e^{-s}_q(t'')\ee{-2isq_\|v_A(t'-t'')}.\label{eq:hteqn}
\end{multline}
Considering long-time  and steady-state behavior (in the variable $t$) with $t',t''>t$, the two-time power
\begin{equation}
  \tilde h^s_k(t,t'')=e^s_k\tgam^s_k(t''-t),
\end{equation}
can be used in Eq.~\eqref{eq:hteqn} to obtain
\begin{multline}
    \pder{}{\tau}\tilde\Gamma^s_k(\tau)=-\epsilon^2\int d\vec pd\vec q\mkpq kpq^2\dkpq kpq e^{-s}_q\\
  \times \int_0^{\tau}d\tau'\tilde\Gamma^s_k(\tau')\ee{-2isq_\|v_A(\tau-\tau')}
\end{multline}
where the one-time spectra $e^s_k$ factored out from the evolution in $t'$. Integrating over variable $\vec p$ and using explicit expression for $\mkpq kpq$, the following equation is obtained for the correlation function $\tgam^s_k(\tau)$ 
\begin{equation}
  \pder{}{\tau}\tgam^s_k(\tau)=-\epsilon^2I(k_\perp)\int_0^{\tau}d\tau'\tgam^s_k(\tau')h_0^{-s}(\tau-\tau')\label{eq:geqn2}
\end{equation}
where
\begin{eqnarray}
  I^s(k_\perp)&=&k_\perp^2\int_0^\infty dq_\perp\mathcal E^{-s}(q_\perp)G(q_\perp/k_\perp)\\
  G(\xi)&\equiv&\int_0^{\pi}\frac{d\phi}{\pi}\frac{(1-\xi\cos\phi)^2\sin^2\phi}{1+\xi^2-2\xi\cos\phi}\\
  h_0^s(\tau)&=&\int_{-\infty}^\infty dk_\|g(k_\|)\ee{2isq_\|v_A\tau}\label{eq:parallelcor}
\end{eqnarray}

Eq.~\eqref{eq:geqn2} is an integro-differential equation describing the scale-dependent correlation for the wave amplitudes $\ppsi kt$ and is the main result of this work. Solutions to this equation, which are discussed below, can then be used to obtain the two-time energy spectrum of MHD turbulence from Eq.~\eqref{eq:hkttp} as
\begin{equation}
    h^s(\vec k,\tau)=e^s(\vec k)\tilde\Gamma^s(\vec k,\tau)\ee{ik_\|v_A\tau}.
\end{equation}

\section{Discussion}
The governing equation for the turbulence correlation function $\tgam^s(\vec k,\tau)$ is an integro-differential equation of the general form
\begin{equation}
\displaystyle\frac{d}{d\tau}\Gamma(\tau)=-\alpha^2\int_0^\tau\Gamma(\tau')h_0(\tau-\tau')d\tau'\label{eq:ide}
\end{equation}
in which $\Gamma,\alpha^2$ and $h_0(\tau)$ play the role of $\tgam^s(\mathbf k,\tau)$, $\epsilon^2I(k_\perp)$ and $ h_0^{-s}(\tau)$, respectively. 
It can be shown that the correlation function $h_0(\tau)$ defined in~\eqref{eq:parallelcor}
\begin{equation}
    h_0^s(\tau)=R^s(r_\|=2v_A\tau)
\end{equation}
results from the substitution of $r_\|=2v_A\tau$ in the spatial correlation function $R^{s}(r_\|)$ between points separated a distance $r_\|$ in the direction of the background magnetic field $\vec B_0$. It then follows that the function $h_0^{s}(\tau)$ can be interpreted as the spatial correlation function of $\vec z^s$ as seen in a frame moving with $\vec z^{-s}$ waves, in which case the speed is $2v_A$. Therefore, the decorrelation time of $h_0^{-s}(\tau)$ corresponds to the time it takes $\vec z^{-s}$ waves to propagate through $\vec z^s$ waves one parallel correlation length, i.e., $\tau_{c\|}\sim 1/2k_\|v_A$.

General solutions to equation \eqref{eq:ide} will be presented and discussed in detail in a separate publication; however, it is worth noting that a closed solution can be obtained to leading order in $\alpha\sim\epsilon$ as follows. First, by formally integrating from time lag $\tau'$ to $\tau$
\begin{equation}
    \Gamma(\tau)-\Gamma(\tau')=-\alpha^2\int_{\tau'}^{\tau}\int_0^{\zeta}\Gamma(\xi)h(\tau-\xi)d\xi d\zeta,
\end{equation}
it is shown that the difference between $\Gamma(\tau)$ and $\Gamma(\tau')$ is of order $\epsilon^2$. As a consequence, replacing $\Gamma(\tau')\simeq\Gamma(\tau)$ inside the integral in Eq.~\eqref{eq:ide} will result only in an $\epsilon^4$ correction, and the equation becomes
\begin{equation}
\displaystyle\frac{d}{d\tau}\Gamma(\tau)=-\alpha^2\Gamma(\tau)H(\tau)
\end{equation}
where $H(\tau)\equiv\int_0^\tau h(\tau-\tau')$. This simplified form of~\eqref{eq:ide} becomes a regular differential equation that can be integrated through elementary methods, subject to the initial condition $\Gamma(0)=1$, namely
\begin{equation}
\Gamma(\tau) = {\rm e}^{-\alpha^2\int_0^\tau H(\tau')d\tau'}.
\end{equation}

A few interesting properties stand out from this solution. First, note that for short times we have to leading order
\begin{equation}
H(\tau) = H(0)+H'(0)\tau+\cdots = h_0(0)\tau\cdots,
\end{equation}
from where it follows
\begin{equation}
\Gamma(\tau) = {\rm e}^{-(\alpha\tau)^2/2}.
\end{equation}

This last approximate solution has the same functional form of previous models of the correlation function in the context of strong MHD turbulence~\cite{narita17a,bourouaine18,bourouaine19,bourouaine20,chhiber19}. A second interesting property of the solution is that because $h_0(\tau)$ is a localized function that vanishes for times much longer than $\tau_{c\|}$, for large $\tau$ we have
\begin{equation}
H(\tau)\approx\lim_{\tau\rightarrow\infty}H(\tau)={\rm const}
\end{equation}
which leads to the asymptotic solution
\begin{equation}
\Gamma(\tau) = {\rm e}^{-\alpha^2 H(\infty)\tau}.
\end{equation}

In summary, for short $\tau$ the solution exhibits Gaussian behavior, while for long time it exhibits exponential behavior. This might explain why exponential decay has been useful in previous analysis of spacecraft and simulation data by a number of authors; see, for instance, Refs.~\cite{matthaeus10,lugones16}.

\section{Conclusion} A weak turbulence closure was obtained for the first time for the two-time energy spectrum of MHD turbulence to leading order in the wave amplitudes. The resulting closure leads to an integro-differential equation for the scale-dependent correlation function $\tilde\Gamma^s(\vec k,\tau)$ in the variable $\tau=t'-t$, which describes the temporal decorrelation of waves with wavevector $\vec k$. The rate of decrease of the correlation at a given scale for $\vec z^s$ waves is determined by the convolution of the correlation function itself with the spatial correlation of $\vec z^{-s}$ waves, measured along the propagation direction, as seen in the frame moving with the $\vec z^s$ waves. 
Approximate solutions were found by assuming the variation of the correlation function is small during the crossing time of counterpropagating waves, exhibiting Gaussian behavior for sufficiently small values of $\tau$ and exponential decay for long time lags $\tau$. These results are largely consistent with similar models of the scale-dependent correlations measured in simulations and in the solar wind, and may find applications in other WT regimes in fluid and plasma turbulence.

\begin{acknowledgements}
This work was supported by grant NNX16AH92G from NASA's Living with a
Star Program and NSF-SHINE grant AGS-1752827.
\end{acknowledgements}
%

\end{document}